\documentclass[journal]{IEEEtran}

\ifCLASSINFOpdf
\else
   \usepackage[dvips]{graphicx}
\fi
\usepackage{url}

\usepackage{graphicx}
\usepackage{graphicx}

\usepackage{latexsym}
\usepackage{amsmath}
\usepackage{amsthm}

\usepackage{amsfonts}
\usepackage{subfigure}
\usepackage{dsfont}
\usepackage{epstopdf}
\usepackage{threeparttable}
\usepackage{multirow}

\usepackage{enumerate}

\usepackage{epsfig}

\ifCLASSINFOpdf
  
\else
 
\fi

\begin{document}

\title{Superimposed Coding Based CSI Feedback Using \\ 1-Bit Compressed Sensing}

\author{Chaojin Qing,~\IEEEmembership{Member,~IEEE}, Qingyao Yang, Bin Cai, Borui Pan and Jiafan Wang
 \thanks{Chaojin Qing, Qingyao Yang and Bin Cai are with School of Electrical Engineering and Electronic Information, Xihua University, Chengdu, China.(e-mail: qingchj@uestc.edu.cn).}
 \thanks{Borui Pan is with Jincheng College of Sichuan University, Chengdu, China (e-mail: pomeloray777@gmail.com).}
 \thanks{Jiafan Wang is with Synopsys Inc., 2025 NE Cornelius Pass Rd, Hillsboro, OR 97124, USA (e-mail: jifanw@gmail.com).}}

 \markboth{IEEE XXX XXX,~Vol.~XX, No.~XX, XXX~2019}%
 {Shell \MakeLowercase{\textit{et al.}}: Bare Demo of IEEEtran.cls for IEEE Journals}

\maketitle

\begin{abstract}
In a frequency division duplex (FDD) massive multiple input multiple output (MIMO) system, the channel state information (CSI) feedback causes a significant bandwidth resource occupation. In order to save the uplink bandwidth resources, a 1-bit compressed sensing (CS)-based CSI feedback method assisted by superimposed coding (SC) is proposed. Using 1-bit CS and SC techniques, the compressed support-set information and downlink CSI (DL-CSI) are superimposed on the uplink user data sequence (UL-US) and fed back to base station (BS). Compared with the SC-based feedback, the analysis and simulation results show that the UL-US's bit error ratio (BER) and the DL-CSI's accuracy can be improved in the proposed method, without using the exclusive uplink bandwidth resources to feed DL-CSI back to BS.
\end{abstract}


\IEEEpeerreviewmaketitle

\section{Introduction}

\IEEEPARstart{A}{s} one of the key technologies for fifth-generation (5G) wireless networks, frequency division duplex (FDD) massive multiple-input multiple-output (MIMO) has drawn increased attention due to the improvement in spectral and energy efficiencies \cite{c1,c2}. With a large number of antennas deployed  at base station (BS), the performance improvement of massive MIMO system significantly relies on accurate channel state information (CSI). In time-division duplexing (TDD) system, the downlink CSI (DL-CSI) can be obtained at BS via the channel reciprocity \cite{c3}. However, the channel reciprocity is not available in FDD massive MIMO system due to the different uplink and downlink spectral bands. Therefore, the DL-CSI should be fed back to base station (BS) through the uplink channel \cite{c4, c5}.

The codebook-based approaches are usually adopted to reduce feedback overhead. Nevertheless, due to the exponential measurement complexity caused by large number of antennas at BS,  this approach is not practical in FDD massive MIMO system \cite{c6}. In \cite{c7}--\cite{c11},  the research results have indicated that many wireless channels have sparse feature. To reduce CSI feedback overhead, compressed sensing (CS)-based CSI feedback scheme is widely used. The approaches exploit the sparsity structures of CSI (e.g., CSI's temporal correlation \cite{c8}, spatial correlation \cite{c9}, and the sparsity-enhancing basis for CSI \cite{c10,c11}) to reduce the channel dimension. Even so, these methods inevitably occupy some uplink bandwidth resources.

In order to avoid CSI feedback independently occupying uplink bandwidth resources, superimposed coding (SC) technique has been introduced into CSI  feedback scheme. \cite{c12} is one of the few works known to us, which combines the SC technology and CSI feedback. In \cite{c12}, after spread processing, the DL-CSI is superimposed on uplink user data sequences (UL-US) and fed back to BS. Although the uplink bandwidth resources have not been occupied, the application of SC results in superposition interference. This degrades the DL-CSI's normalized mean squared error (NMSE) and the UL-US's bit error ratio (BER).

To improve the DL-CSI's NMSE and the UL-US's BER, this paper combines 1-bit CS  \cite{c13}, SC technique and support-set feedback method. In practice, 1-bit quantization is particularly attractive because the construction of the quantizer is simple and cost-effective  \cite{c14, c14a}. To the best of our knowledge, the SC-based CSI feedback using 1-bit CS method for FDD massive MIMO systems has not been studied in existing literatures. The main contributions of this paper are summarized as follows:
\begin{enumerate}
  \item Introducing ``1-bit CS'' technique into the SC-based CSI feedback scheme improves DL-CSI's NMSE and UL-US's BER. In \cite{c12}, an unquantized and uncoded DL-CSI is estimated, and then the estimated DL-CSI is used to reduce superimposed interference. Unlike the case in \cite{c12}, 1-bit CS transforms a CSI estimation problem into the problem of bit (or sign) information detection and then 1-bit CS reconstruction. Since only the bit (or sign) information needs to be detected, the interference cancellation is more effective than that of \cite{c12}. Thus, both UL-US's BER and DL-CSI's NMSE can be improved.
  \item The support-set of CSI is superimposed on UL-US and fed back to BS to further improve the NMSE of DL-CSI. In CS-based CSI feedback schemes, the support-set of CSI is required to be recovered at BS \cite{c15}, \cite{c18}. Without recovering the support-set, the number of measurements could be significantly reduced \cite{c18} and the superimposed data is sharply reduced. And then, the spread spectrum gain of SC-based CSI feedback is effectively improved.
  \item From \cite{c19}, the accuracy of reconstruction algorithm can be effectively improved with the priori information of support-set. Based on the de-spread support-set and binary iterative hard thresholding (BIHT) algorithm \cite{c20} (other similar reconstruction algorithms can be applied as well), a SC-aided BIHT (SCA-BIHT) algorithm is proposed to improve the recovery of the compressed DL-CSI at BS.
\end{enumerate}

\textit{Notation}: Boldface letters are used to denote matrices and column vectors; $(\cdot)^{T}$, $(\cdot)^{H}$ and $(\cdot)^{-1}$ denote the transpose, conjugate transpose, matrix inversion. $\mathbf{I}_p$ is the identity matrix of size $P\times P$, $\mathbf{0}$ is the matrix or vector with all zero elements, the $l_2$ norm of a vector $\mathbf{x}$ is written as $\lVert\mathbf{x}\rVert_2$. $\odot$ denotes the operation of Hadamard product for two vectors or matrices. $\eta_k(\mathbf{x})$ represents computing the best $k$-term approximation of $\mathbf{x}$ by thresholding. $\mathrm{dec}(\cdot)$ is the hard decision operation, in which the current data is determined as the modulated data with the smallest Euclidean distance from current data. $sgn(\cdot)$ denotes an operator that performs the sign function element-wise on the vector, e.g., the sign function returns  $+1$  for positive numbers and $(0)$ otherwise.

\section{System model}

We consider a massive MIMO system consists of a BS with $N$ antennas and $K$ single-antenna users. The DL-CSI is superimposed on UL-US and then fed back to BS. In this way, the overhead of uplink bandwidth resources, particularly used to feed back DL-CSI, is avoided. Similar to  \cite{c8}--\cite{c11}, we assume the DL-CSI has been estimated at users and mainly focus on  CSI feedback\footnote{Note that, due to limited computational power, the users should employ some low-complexity channel estimation methods, which require consideration but go beyond the scope of this letter. Since we mainly focus on CSI feedback, we assume that the DL-CSI has been estimated perfectly at user.}. After the processing of matched-filter (MF) (i.e., the conventional multiuser detector structure consists of a MF bank front \cite{c20_1}), the received signal $\widetilde{\mathbf{Y}}_k$ sent by the $k$-th user, $k=1,2,\dots,K$ can be given by
\begin{equation}
\begin{aligned}
\label{equ:received_signal_Y}
\mathbf{\widetilde{Y}}_{k} =& \mathbf{G}\mathbf{X}  + \widetilde{\mathbf{n}}_k
\\ =& \mathbf{G}(\sqrt{\rho E_{k}}\mathbf{s}_{k} + \sqrt{(1-\rho)E_{k}}\mathbf{d}_{k}) + \widetilde{\mathbf{n}}_k ,
%\widetilde{\mathbf{Y}_{k}}=& \mathbf{G}\mathbf{X}_{k}  + \widetilde{\mathbf{N}}_k\\
%=& \mathbf{G}(\sqrt{\rho E_{k}}\mathbf{s}_{k} + \sqrt{(1-\rho)E_{k}}\mathbf{d}_{k}) + \widetilde{\mathbf{N}}_k ,
\end{aligned}
\end{equation}
where $\mathbf{G}$ is a $N\times1$ uplink channel matrix, $\rho \in [0,1]$ stands for the power proportional coefficient of DL-CSI, $E_{k}$ represents the total transmitting power, $\mathbf{d}_k \in \mathbb{C}^{1\times P}$ denotes the UL-US signal; $\widetilde{\mathbf{n}}_k \in \mathbb{C}^{N\times P}$ represents the feedback link noise whose elements are with zero-mean and variance $\sigma^2_n$ \cite{c6}; In particular, the $\mathbf{s}_k \in \mathbb{C}^{1\times P}$ in (\ref{equ:received_signal_Y}) is the superposition signal that consists of compressive DL-CSI, sparsity and support-set. III-A would expatiate the $\mathbf{s}_k$ to this paper.

\section{The Proposed Feedback Method Using 1-Bit CS}
\label{sec:guidelines}

In this section, we first present how to introduce 1-bit CS technique into the SC-based CSI feedback scheme (see III-A). Then, in III-B, the DL-CSI reconstruction and UL-US detection are described, where we especially explain the details of proposed SCA-BIHT. Finally, the analysis of computational complexity for SCA-BIHT is given.

\subsection{SC based DL-CSI Feedback}
After exploiting the sparsity structure (by using methods mentioned in \cite{c7}--\cite{c11}), the sparse DL-CSI ${\mathbf{h}}_{k}$ can be compressed according to 1-bit CS technique, i.e.,
\begin{equation}
\label{equ2}
\left\{ \begin{array}{l}
{{\mathbf{y}}_{real}} = sgn\left( {{\mathop{\mathrm {Re}}\nolimits} \left( {{{{\mathbf{ h}}}_{k}}{{\mathbf{\Phi }}_k}} \right)} \right)\\
{{\mathbf{y}}_{imag}} = sgn\left( {{\mathop{\mathrm {Im}}\nolimits} \left( {{{{\mathbf{ h}}}_{k}}{{\mathbf{\Phi }}_k}} \right)} \right)
\end{array},\right.
\end{equation}
where $\mathbf{\Phi}_{k} $ is a $N \times M$ measurement matrix and the DL-CSI $\mathbf{{h}}_{k}$ is a ${1\times N}$ vector. In (\ref{equ2}), $\mathbf{y}_{real}$ and $\mathbf{y}_{imag}$ are used to represent the DL-CSI compression's real and imaginary parts, respectively.

We assume the  DL-CSI ${\mathbf{h}}_{k}$ features $\xi_k$-sparsity, i.e., only $\xi_k$ non-zero elements in ${\mathbf{h}}_{k}$ \cite{c7}. For convenience, a set ${\mathbf{z}}_k \in \{ 0,1 \} ^{1\times N}$ is employed to label the set of indices of DL-CSI's non-zero elements (i.e., the support-set). That is, the index of DL-CSI's zero elements is labeled by 0 and the index of DL-CSI's non-zero elements is labeled by 1. For example, ${\mathbf{h}}_{k} = (h_1,h_2,\dots,h_5)$ and $\mathbf{z}_k= [1,1,0,0,1]$ mean the value of $h_1$, $h_2$ and $h_5$ are non-zero elements, $h_3=h_4=0$.

With the bit-form of $\xi_k$ as $\mathbf{k}_{bin} \in \{ 0,1 \} ^{B}$, the feedback vector $\mathbf{w}_{k}$, which merges $\mathbf{y}_{real}$, $\mathbf{y}_{imag}$, $\mathbf{z}_k$ and $\mathbf{k}_{bin}$, can be expressed as
\begin{equation}
\label{equ:unmodulated_signal_w}
\mathbf{w}_{k} = [\mathbf{y}_{real},\mathbf{y}_{imag},\mathbf{z}_{k},\mathbf{k}_{bin}].
\end{equation}
It is worth noting that the elements of $\mathbf{w}_{k}$ only contain 0 and 1, which can be viewed as a bit stream. With digital modulation, such as the quadrature phase shift keying (QPSK), $\mathbf{w}_{k}$ is mapped to a $1\times L$ modulated vector $\mathbf{x}_{k}$. Without loss of generality, the UL-US's length $P$ is larger than $L$ due to main task of the user services. Thus, a spreading method can be utilized to capture a spread spectrum gain. The superposition signal $\mathbf{s}_{k}$  can be obtained via the using of pseudo-random codes (e.g., the Walsh codes) to spread $\mathbf{x}_{k}$, i.e.,

\begin{equation}
\label{equ:received_signal_L}
\mathbf{s}_{k} = \mathbf{x}_{k}\mathbf{q}^{{T}},
\end{equation}
where $\mathbf{q} \in\mathbb{R}^{ P\times L}$ consists of $L$ codes of length $P$ satisfying $\mathbf{q}^{T}\mathbf{q}=P\cdot\mathbf{I}_L$. Then the superposition signal $\mathbf{s}_{k}$ and UL-US $\mathbf{d}_k$ are weighted and superimposed, i.e., $\sqrt{\rho E_{k}}\mathbf{s}_{k} + \sqrt{(1-\rho)E_{k}}\mathbf{d}_{k}$,

and fed back to BS, which is described in (\ref{equ:received_signal_Y}) as well.

%=============== TABEL I  ============================
\begin{table}[t]\normalsize
\caption{Sca-biht Algorithm}\normalsize
\label{table}
\setlength{\tabcolsep}{3pt}
\begin{tabular}{l}
\hline

\textbf{Input}: measurement matrix $\mathbf{\Phi}_{k}$, real part and imaginary part \\
 \kern 8pt of 1-bit noise measurement ($\widetilde{\mathbf{y}}_{real}$ and $\widetilde{\mathbf{y}}_{imag}$), sparsity $\widetilde \xi_k$, \\
  \kern 8pt and received support-set $\widetilde{\mathbf{z}}_k$. \\

\textbf{Initialize:} maximum number of iterations $Itermax$, iteration \\
    \kern 8pt count $t=0$, the real part and imaginary part of reconstruct-\\
        \kern 8pt  ed data are set to zero, i.e., $\mathbf{r}^0_{real}=\textbf{0}$ and $\mathbf{r}^0_{imag}=\textbf{0}$.\\
\textbf{Begin}: \\
    \kern 8pt $1)$ Increment: $t=t+1$;\\

    \kern 8pt $2)$ Gradient update: \\
        \kern 20pt     $\mathbf{r}^t_{real} = \eta_{\widetilde{\xi}_k}(\mathbf{r}^{t-1}_{real}+ (\widetilde{\mathbf{y}}_{real} - sgn(\mathbf{r}^{t-1}_{real}\mathbf{\Phi}_{k}))\mathbf{\Phi}_{k}^{T})$, \\
            \kern 20pt    $\mathbf{r}^t_{imag}=\eta_{\widetilde{\xi}_k}(\mathbf{r}^{t-1}_{imag}+ (\widetilde{\mathbf{y}}_{imag} - sgn(\mathbf{r}^{t-1}_{imag}\mathbf{\Phi}_{k}))\mathbf{\Phi}_{k}^{T})$,\\
    \kern 20pt $\mathrm{supp}(\mathbf{r}^{t}) = \mathrm{supp}(\mathbf{r}^t_{real})\cup \mathrm{supp}(\mathbf{r}^t_{imag})$; \\
    \kern 8pt $3)$ Go to step 5) if $\mathrm{supp}(\mathbf{r}^{t})\cap \widetilde{\mathbf{z}}_k = \emptyset$, else\\
    \kern 20pt go to the next step;\\
    \kern 8pt $4)$ Auxiliary correction:\\

         \kern 20pt    $\mathbf{r}^t_{real} = \mathbf{r}^t_{real} \odot \widetilde{\mathbf{z}}_k$, \\
         \kern 20pt    $\mathbf{r}^t_{imag} = \mathbf{r}^t_{imag} \odot \widetilde{\mathbf{z}}_k$;\\

    \kern 8pt $5)$ Go to step $1)$ if $t<Itermax$, else go to next step;\\
    \kern 8pt $6)$ Combination:\\
         \kern 20pt    $\widetilde{\mathbf{H}}=\mathbf{r}^t_{real} + i \times \mathbf{r}^t_{imag}$;\\
    \kern 8pt $7)$ Normalization:\\
         \kern 20pt    $\widehat{\mathbf{h}}_{k} = \widetilde{\mathbf{H}}/\lVert\widetilde{\mathbf{H}}\rVert_2$;\\
\textbf{End} \\
\textbf{Output}: Reconstructed DL-CSI $\widehat{\mathbf{h}}_{k}$.\\

\hline
\end{tabular}
\end{table}

%================table2===============

\begin{table}[!t]

\renewcommand{\arraystretch}{1.3}
\caption{Computational Complexity}
\label{table2}
\centering
\begin{tabular}{|c|c|}
\hline
Algorithm&Complexity\\
\hline
BIHT & $\mathcal{O}((MN)*Iter1)$\\
\hline
SCA-BIHT & $\mathcal{O}((MN)*Iter2)$\\
\hline
\end{tabular}
\end{table}

%===============Fig 1 (a) and (B)  ============================
\begin{center}
\begin{figure*}[!hbtp]
\centering
  \subfigure[BER vs. SNR with the same bit-overhead, where $\rho=0.2$.]{
    \label{figure1_1bit_BER} %% label for first subfigure
    \includegraphics[width=0.8\columnwidth]{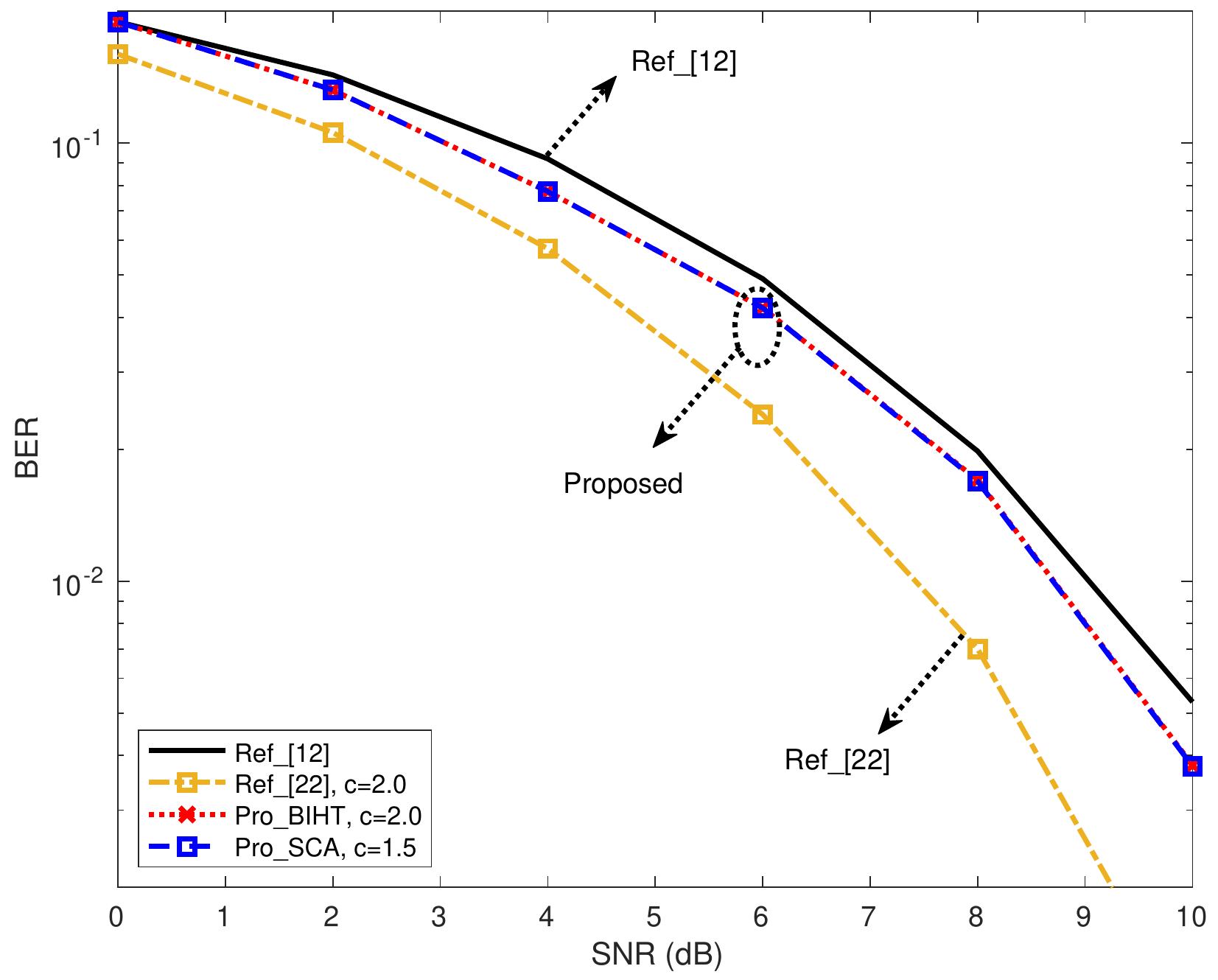}}
  \subfigure[NMSE vs. SNR with the same bit-overhead, where $\rho=0.2$.]{
    \label{figure2_1bit_mse} %% label for second subfigure
    \includegraphics[width=0.8\columnwidth]{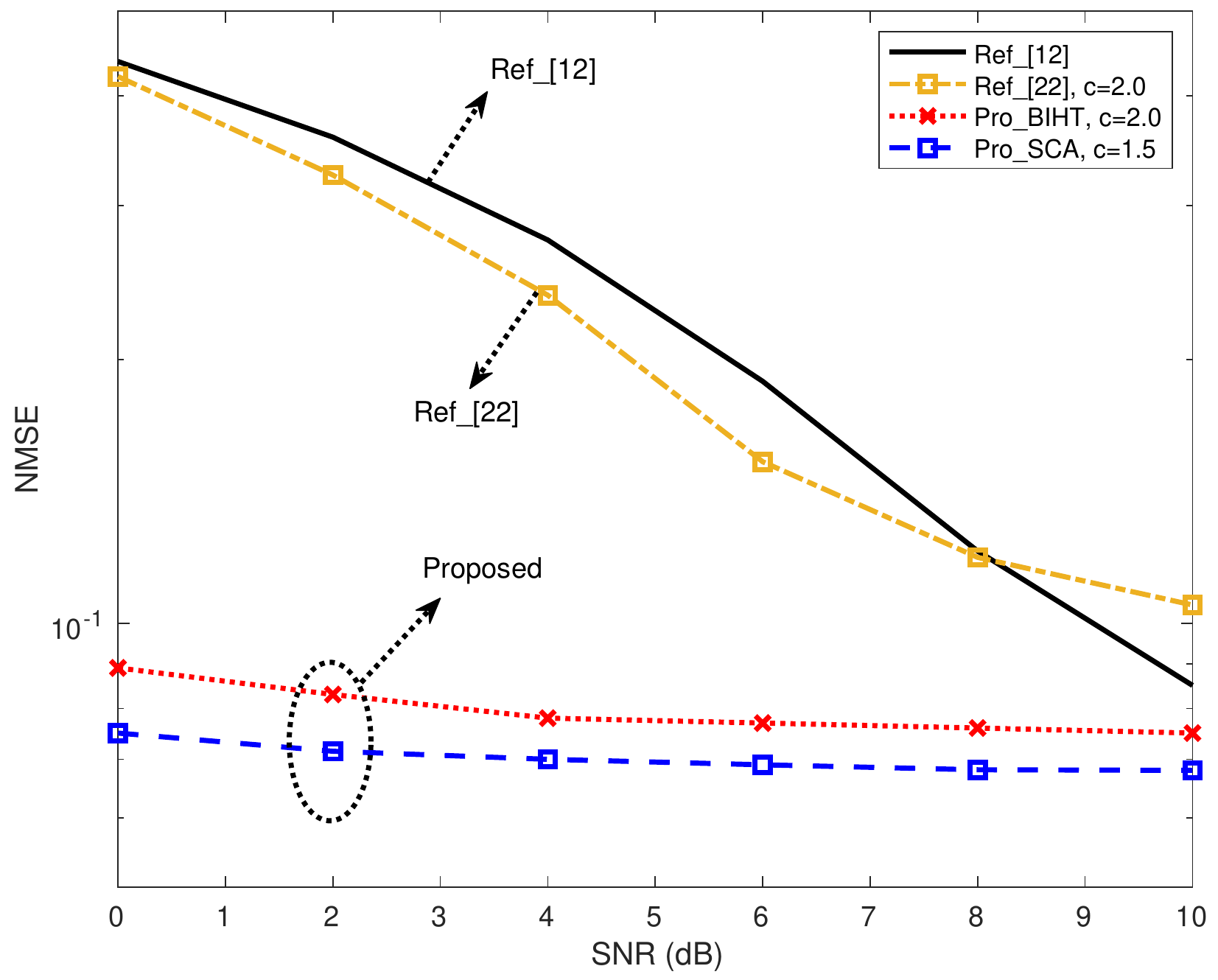}}
\caption{The BER and NMSE of different schemes.}
%performance comparisons of BER and NMSE, where $\rho=0.2$.}
%compression ratios of ``Prop-BIHT'' and ``Prop-SCA'' are set to $c=2$ and $c=1.5$, respectively.}$N=64$, $P=1024$,
% \captionsetup{justification=centering}
\label{1bit}
\end{figure*}
\end{center}

\subsection{UL-US Detection and DL-CSI Reconstruction}
\subsubsection{UL-US Detection}
With the received signal $\widetilde{\mathbf{Y}}_k$ in equation~(\ref{equ:received_signal_Y}), the de-spread signal can be obtained by
\begin{equation}
\begin{aligned}
\label{equ:depread_signal_Z}
\widetilde{\mathbf{x} }=& \frac{1}{P}\widetilde{\mathbf{Y}}_{k}\mathbf{q}\\
=& \sqrt{\rho E_k}\mathbf{G}\mathbf{x}_{k}+\frac{1}{P}\sqrt{(1-\rho)E_k}\mathbf{G}\mathbf{d}_{k}\mathbf{q}+\frac{1}{P}\widetilde{\mathbf{n}}_{k} \mathbf{q}.
\end{aligned}
\end{equation}
Subsequently, the estimation that contains DL-CSI and support-set can be acquired via minimum mean square error (MMSE) detection, i.e.,
\begin{equation}
\label{equ:MMSE_detection_r}
\begin{array}{l}
{{{\bf{\widetilde x}}}_{\mathrm{MMSE}}} = \mathrm{dec}\left( {P\sqrt {\rho {E_k}} } \right.\left[ {\left( {1 + \left( {P - 1} \right)\rho } \right){E_k}{{\bf{G}}^{H}}{\bf{G}}} \right.\\
\kern 43pt {\left. { + ~{\sigma^2_n}} \right]^{ - 1}}{{\bf{G}}^{H}}{\bf{\widetilde x}}\left. {} \right).
\end{array}
\end{equation}
Then, taking advantage of interference cancellation described in \cite{c12}, the interference caused by DL-CSI can be eliminated in such a way:
\begin{equation}\small
\label{equ:UL_US_received_signal_sk}
\begin{aligned}
{{{\bf{\widehat d}}}_k} =&
{{{\bf{\widetilde Y}}}_k} - \sqrt {\rho {E_k}} {\bf{G}}{{{\bf{\widetilde x}}}_{\mathrm{MMSE}}}{{\bf{q}}^{T}}
\kern 0pt \\= &\sqrt {\left( {1 - \rho } \right){E_k}} {\bf{G}}{{\bf{d}}_k}+\sqrt {\rho {E_k}} {\bf{G}}\left( {{{\bf{x}}_k} - {{{\bf{\widetilde x}}}_{\mathrm{MMSE}}}} \right){{\bf{q}}^{T}} + \widetilde{\mathbf{n}}_{k}.
\end{aligned}
\end{equation}
With the application of MMSE detection in $ (\ref{equ:UL_US_received_signal_sk})$, the estimated UL-US $ {\bf{\widehat d}}_k$ is obtained, then $\mathbf{w}_k$ can be recovered from $\widetilde{\mathbf{x}}_{\mathrm {MMSE}}$. The sign information $\widetilde{\mathbf{y}}_{real}$, $\widetilde{\mathbf{y}}_{imag}$, sparsity $\widetilde \xi_k$, and support-set $\widetilde{\mathbf{z}}_k$  can be restored via the position relation in equation~(\ref{equ:unmodulated_signal_w}).

\subsubsection{DL-CSI Reconstruction}
BS can recover the DL-CSI via the SCA-BIHT algorithm, with the recovered sign information $\widetilde{\mathbf{y}}_{real}$, $\widetilde{\mathbf{y}}_{imag}$, sparsity $\widetilde{\xi}_k$, and support-set $\widetilde{\mathbf{z}}_{k}$. The details of SCA-BIHT are shown in \textbf{TABLE~\ref{table}}. Similar to \cite{c14} \cite{c24}, the direction of the reconstructed signal is obtained via the normalization step, i.e., step 7) in SCA-BIHT. Need to mention that, we propose SCA-BIHT to improve BIHT, and other similar reconstruction algorithms can naturally be improved according to the same approach. In SCA-BIHT, the input and auxiliary correction are different from BIHT, which are described as follows:
\begin{itemize}
  \item \textbf{Input and Initialization of SCA-BIHT}: The input includes the received support-set $\widetilde{\mathbf{z}}_{k}$, which is not contained in BIHT \cite{c20}. Since BS does not need to reconstruct support-set, the proposed method has fewer iterations and lower computational complexity (see III-C for details).

  \item \textbf{Auxiliary Correction}: As shown in step 4), the reconstructed values are corrected by using the received support-set $\widetilde{\mathbf{z}}_{k}$. That is, according to the position of 0 elements in $\widetilde{\mathbf{z}}_{k}$, the elements at the corresponding position in the reconstructed value are set to 0, and the remaining elements are unchanged. But the BIHT doesn't contain support-set correction.

 \end{itemize}

  Compared with the BIHT, the proposed SCA-BIHT is more concise, due to the auxiliary of support-set.

\subsection{Computational Complexity}
 The comparison of computational complexity between \text{BIHT} and \text{SCA-BIHT} is given in \text{TABLE~\ref{table2}}, where $Iter1$ and $Iter2$ denote the iteration number of \text{BIHT} and \text{SCA-BIHT}, respectively. For each iteration, SCA-BIHT and BIHT have the computational complexity $\mathcal{O}(MN)$. Despite all this, SCA-BIHT has fewer iterations than BIHT, i.e., $Iter2< Iter1$, due to no requirement of support-set reconstruction. Thus, SCA-BIHT has lower computational complexity than that of BIHT.

%===============Fig 2 (a) and (B)  ============================
\begin{center}
\begin{figure*}[!hbtp]
\centering
    \subfigure[The impacts of different $\rho$ and $c$ on BER of Prop-SCA.]{
    \label{figure3_support_set_ber}
    \includegraphics[width=0.8\columnwidth]{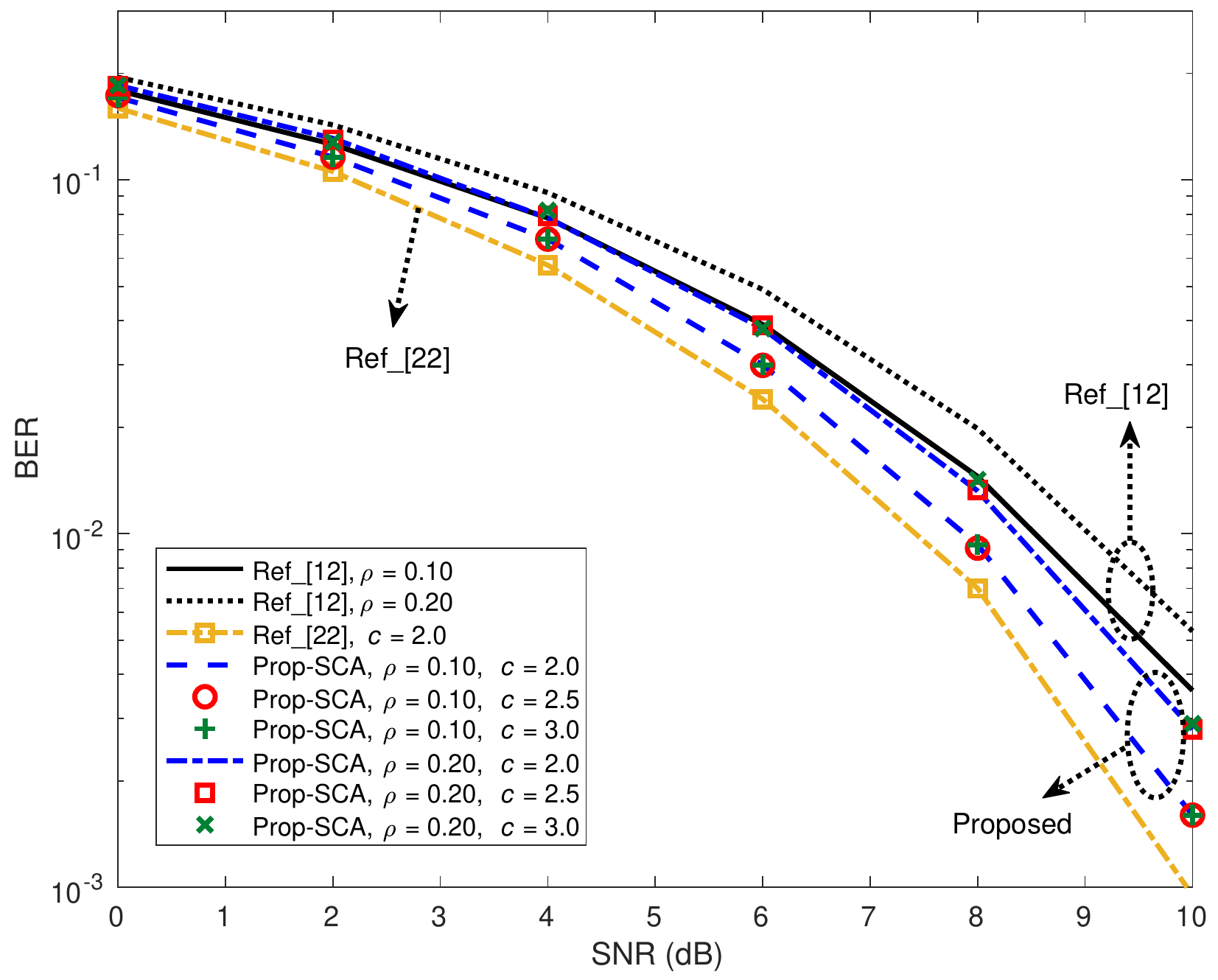}}
    \subfigure[ The impacts of different $\rho$ and $c$ on NMSE of Prop-SCA.]{
    \label{figure4_support_set_mse}
    \includegraphics[width=0.8\columnwidth]{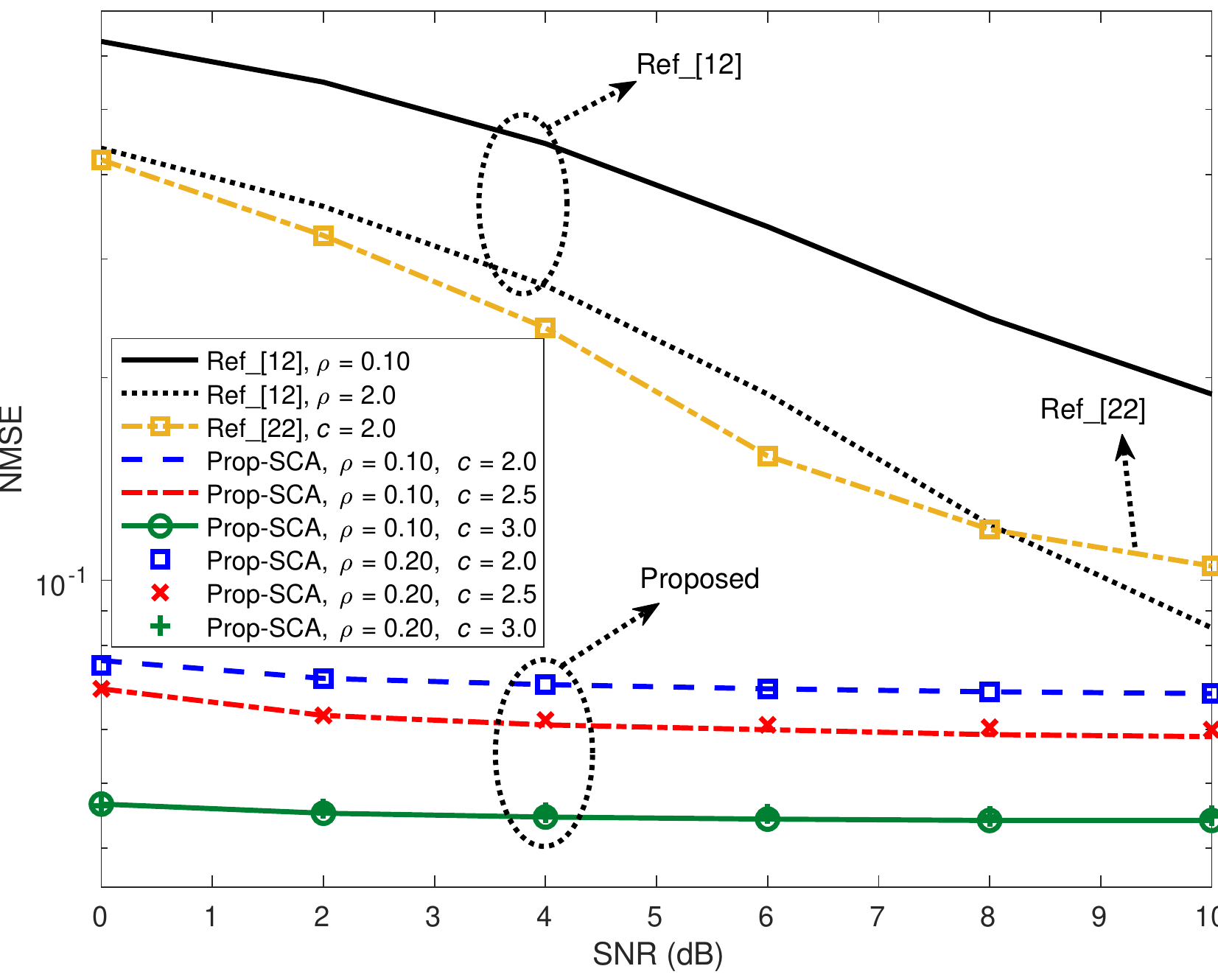}}
     \caption{Performance of Prop-SCA for various $\rho$ and $c$.}

\label{support-set}
\end{figure*}
\end{center}

\section{Experiment results}
In this section, we give some numerical results of the SC based CSI feedback with 1-bit CS under different conditions. The basic parameters involved are listed below. ${\mathbf{h}}_{k}$ features $\xi_k$-sparsity, whose elements obey $\mathcal{CN}(0,1)$. The $N\times M$ measurement-matrix $\mathbf{\Phi}_{k}$ is set as a Gaussian random matrix, whose elements obey $\mathcal{N}(0,1)$ \cite{c20} \cite{c24}. UL-US is a $1\times P$ complex sequence modulated by quadrature phase shift keying (QPSK). We set $P=1024$, $N=64$, $\xi_k=8$, and $Itermax=100$. The sampling rate $c$, signal-to-noise ratio (SNR) in decibel (dB), and NMSE are defined as ${c}={M}/{N}$, $\text{SNR} = 10\textrm{log}_{10}(E_k/\sigma^2_n)$ and $\text{NMSE} = {{\left\| {{{\bf{h}}_k} - {{{\bf{\widehat h}}}_k}} \right\|_2^2} \mathord{\left/
 {\vphantom {{\left\| {{{\bf{h}}_k} - {{{\bf{\hat h}}}_k}} \right\|_2^2} {\left\| {{{\bf{h}}_k}} \right\|_2^2}}} \right.
 \kern-\nulldelimiterspace} {\left\| {{{\bf{h}}_k}} \right\|_2^2}}$, respectively. Three iterations of interference cancellation are employed for \cite{c12}, while only one iteration for the proposed scheme. In \cite{c12}, simulations show that with three iterations, the SC-based feedback algorithm nearly converges. According to (\ref{equ:depread_signal_Z})--(\ref{equ:UL_US_received_signal_sk}) and  DL-CSI reconstruct algorithm in TABLE~\ref{table}, the interference cancellation in proposed scheme is performed only one time. More iterations could not obtain significant improvement but merely increase the complexity. According to \cite{c25}, a feedback method based on 1-bit CS is also employed for time division multiplexing (TDM) mode in our experiments, where $P$ modulated UL-US and $cN$ modulated DL-CSI are up-transmitted with an additional $12.5\%$ of uplinking bandwidth are occupied, i.e., $cN/P=12.5\%$ with $c=2$.

 For simplicity, ``Prop-SCA'' is used to denote the proposed SC-based CSI feedback; ``Prop-BIHT'' represents the SC-based CSI feedback without support-set  ${\mathbf{z}}_{k}$, and BS adopts BIHT for DL-CSI reconstruction;  ``Ref\_[12]'' denotes the SC method in \cite{c12};  ``Ref\_[22]'' denotes the feedback method in \cite{c25} with TDM mode, i.e., the TMD-based feedback.

To verify the effectiveness of proposed scheme. We first make the performance comparison between Ref\_[12],  Ref\_[22], Prop-BIHT and Prop-SCA  in Fig.~\ref{1bit}. Where $\rho = 0.2$, the sampling rates of scheme Ref\_[22], Prop-BIHT and Prop-SCA are respectively set as $c=2$, $c=2$ and $c=1.5$. It is worth noting that this parameter setting of sampling rate is employed to promote Ref\_[22], Prop-BIHT and Prop-SCA have the same bit-overhead to bear CSI feedback. For Prop-BIHT and Ref\_[22], the bit-overhead is $256$ bits according to $c\times N \times 2 =2\times 64 \times 2 = 256$, where we product 2 is due to the consideration of real and imaginary parts. The same bit-overhead can be obtained in Prop-SCA by computing $c\times N \times 2 +N = 1.5\times 64 \times2 +64 = 256$ bits, where we add $N$ is due to the bit-overhead of support-set feedback.

Fig.~\ref{figure1_1bit_BER} shows that the BER performance of the proposed scheme is better than Ref\_[12] and worse than  Ref\_[22]. The interference cancellations in Prop-BIHT and Prop-SCA are effective than that of Ref\_[12] due to the introducing of 1-bit CS. Since the same modulation, power proportional coefficient and SC model are utilized, the identical bit-overhead brings UL-US the equal superimposed interference from DL-CSI (see (\ref{equ:received_signal_Y})). Thus, the similar BERs of Prop-BIHT and Prop-SCA are observed in Fig.~\ref{figure1_1bit_BER}.
Without any superimposed interference, Ref\_[22] obtains lower BER than that of superimposition modes (i.e., Ref\_[12], Prop-BIHT and Prop-SCA) at the cost of additional 12.5\% uplink bandwidth resources.

In Fig.~\ref{figure2_1bit_mse}, the NMSEs of proposed Prop-BIHT and Prop-SCA are much smaller than that of Ref\_[12] and Ref\_[22], where the parameter settings are the same as those in Fig.~\ref{figure1_1bit_BER}. The NMSE of Prop-BIHT is no more than 0.08 in the entire SNR, while the much larger NMSEs are encountered by Ref\_[12] and Ref\_[22], e.g., $0.274$ and $0.237$ for Ref\_[12] and Ref\_[22], respectively, when $\mathrm {SNR} = 4$dB. Obviously, the proposed Prop-BIHT and Prop-SCA improve the NMSEs of Ref\_[12] and Ref\_[22]. Furthermore, it can be observed that, Prop-SCA can further improve the NMSE performance of Prop-BIHT. In Fig. 1(b), Prop-SCA reaches smallest NMSE, which is clearly lower than that of Prop-BIHT. With the same bit-overhead as Prop-BIHT, Prop-SCA further improves the NMSE of DL-CSI by feeding support-set back to BS.

From Fig.~\ref{figure1_1bit_BER} and Fig.~\ref{figure2_1bit_mse}, introducing 1-bit CS technology into SC-based CSI feedback can improve the DL-CSI's NMSE and UL-US's BER. Furthermore, due to the increase of spread spectrum gain, the support-set feedback promotes the proposed Prop-SCA further improve the NMSE of Prop-BIHT. Compared to Ref\_[22] (i.e., TMD-based feedback), although the BERs are sacrificed due to the superimposed interference, the 12.5\% uplink bandwidth savings and much lower NMSE are captured by of Prop-BIHT and Prop-SCA.

To demonstrate the impacts of different $\rho$ and $c$ on Prop-SCA, the BER and NMSE performances are respectively given in Fig.~\ref{figure3_support_set_ber} and Fig.~\ref{figure4_support_set_mse}, where different $\rho$ (i.e., $\rho=0.1$ and $\rho=0.2$) and different $c$ (i.e., $c=2.0$, $c=2.5$, and $c=3.0$) are considered.

Fig.~\ref{figure3_support_set_ber} illustrates the BER performance of Prop-SCA with SNR varying from $0$dB to $10$dB. It is obvious that the Prop-SCA evidently improves the BER when compared to Ref\_[12] with the equal $\rho$, especially for a relatively high SNR, e.g., $\mathrm{SNR}>2$dB. For each $\rho$, the impact of $c$ on BERs of Prop-SCA and Ref\_[12] is not clear, that is because the identical bit-overhead (superimposed on UL-US with the same length $P=1024$) makes the equal superimposed interference (from DL-CSI) encountered by UL-US. On the whole, compared to Ref\_[12], Fig.~\ref{figure3_support_set_ber} shows that the BER improvement of Prop-SCA possesses a good robustness against the impacts of $\rho$ and $c$. In addition, a smaller $\rho$ narrows the BER gap between Prop-SCA and Ref\_[22] due to a weaker superimposed interference (encountered by Prop-SCA).

 To validate the robustness of NMSE against the impact of $\rho$ and $c$ on Prop-SCA, the NMSE performance is given in Fig.~\ref{figure4_support_set_mse}. This figure reflects that, compared with Ref\_[12] and Ref\_[22], Prop-SCA obtains smaller NMSEs. As $c$ increases, the NMSE of Prop-SCA can be improved due to the increase of measurements, and not significantly affected by the change of $\rho$. The reason is that the Prop-SCA transforms a CSI estimation problem into the sign detection problem. With the using of SCA-BIHT in TABLE~\ref{table}, the detected noise measurements (i.e., $\widetilde{\mathbf{y}}_{real}$ and $\widetilde{\mathbf{y}}_{imag}$) leads to less obvious reconstruction differences of DL-CSI with the various $\rho$.

To sum up, compared to SC-based feedback, the proposed Prop-BIHT and Prop-SCA can improve the UL-US's BER and the DL-CSI's NMSE. Compared to Prop-BIHT, the Prop-SCA can further improve the DL-CSI's NMSE performance. Without using the exclusive uplink bandwidth resources, the Prop-SCA can improve the DL-CSI's NMSE performance of TMD-based feedback (i.e., Ref\_[22]). A small $\rho$ (e.g., $\rho = 0.1$) guarantees the UL-US's BER of Prop-SCA is only slightly degraded relative to Ref\_[22], while saving 12.5\% uplink bandwidth resources and keeping improvement of DL-CSI's NMSE. In addition, the Prop-SCA possesses a good robustness against the impact of $\rho$ and $c$. Thus, introducing ``1-bit CS'' technique into SC-based CSI feedback brings us great benefits, and the support-set feedback is attractive.

\section{Conclusion}
In the proposed method, SC technique avoids the occupation of uplink bandwidth resources, 1-bit CS method transforms the DL-CSI estimation problem into a bit (or sign) information detection problem, feeding support-set back to BS significantly reduces the superimposed data, and then the interference cancellation and spread spectrum gain can be effectively improved. Meanwhile, proposed method also adopts SCA-BIHT algorithm to reconstruct DL-CSI at BS as well. The analysis and simulation results show that the proposed method can improve the UL-US's BER and the DL-CSI's NMSE, compared with traditional SC-based DL-CSI feedback method. Although the UL-US's BER is affected by the application of SC, a relatively small power proportional coefficient can still guarantee the BER performance of the proposed method is only slightly degraded relative to TMD-based feedback, while significantly saving uplink bandwidth resources and improving DL-CSI's NMSE.

\appendices

\ifCLASSOPTIONcaptionsoff
  \newpage
\fi


\begin{thebibliography}{00}



\bibitem{c1} Y. Li, C. Tao, A. Swindlehurst, A. Mezghani and L. Liu, ``Downlink achievable rate analysis in massive MIMO systems with one-bit DACs'', \emph{IEEE Commun. Lett.}, vol. 21, no. 7, pp. 6469-6472, Jul. 2017.
\bibitem{c2}  L. Nguyen, T. Duong, H. Ngo and K. Tourki, ``Energy efficiency in cell-free massive MIMO with zero-forcing precoding design'', \emph{IEEE Commun. Lett.}, vol. 21, no. 8, pp. 1871-1874, Aug. 2017.
\bibitem{c3} J. Mirza, B. Ali, S. S. Naqvi and S. Saleem, ``Hybrid precoding via successive refinement for millimeter wave MIMO communication systems'', \emph {IEEE Commun. Lett.}, vol. 21, no. 5, pp. 991-994, Jan. 2017.
\bibitem{c4} X. Luo, P. Cai, C. Shen, D. Hu, and H. Qian, “DL-CSI acquisition and feedback in FDD massive MIMO via path aligning”, in \emph {ICUFN., Milan}, Italy, Jul. 2017, pp. 349–354.

\bibitem{c5} R. Zhang, H. Zhao, and J. Zhang, ``Distributed compressed sensing aided sparse channel estimation in FDD massive MIMO system'', \emph{IEEE Access}, vol. 6, pp. 18383--18397, Mar. 2018.
\bibitem{c6} T. Jiang, M. Song, X. Zhao, and X. Liu, ``A codebook-adaptive feedback algorithm for cellular-based positioning'', \emph{IEEE Access.}, vol. 6, pp. 32109--32164, Jun. 2018.
\bibitem{c7} Z. Lv, Y. Li, ``A channel state information feedback algorithm for massive MIMO systems'', \emph{IEEE Commun. Lett.}, vol. 20, no. 7, pp. 1461-1464, Jul. 2016.
\bibitem{c8} W. Shen, L. Dai, Y. Shi, X. Zhu, and Z. Wang, ``Compressive sensing-based differential channel feedback for massive MIMO'', \emph{ Electron Lett.}, vol. 51, no. 22, pp. 1824--1826, Oct. 2015.
\bibitem{c9} Y. Liao, X. Yang, H. Yao, et al. "Spatial correlation based channel compression feedback algorithm for massive MIMO systems", \emph{Digital Signal Processing }, 2019.
\bibitem{c10}L. Lu, Y. Li, D. Qiao, and W. Han, ``Sparsity-enhancing basis for compressive sensing based channel feedback in massive MIMO systems'', \emph { Proc. IEEE Int. Conf. Global Commun}., San Diego, CA, Dec. 2015, pp. 1--6.
\bibitem{c11} M. S. Sim, J. Park, C.-B. Chae, R. W. Heath, "Compressed channel feedback for correlated massive MIMO systems", \emph { Proc. IEEE Int. Conf. Commun. (ICC)}, pp. 360-364, Jun. 2014.
\bibitem{c12} D. Xu, Y. Huang, and L. Yang, ``Feedback of downlink channel state information based on superimposed coding'', \emph {IEEE Commun. Lett.}, vol. 11, no. 3, pp. 240--242, Mar. 2007.

\bibitem{c13} P. Boufounos and R. Baraniuk, ``1-bit compressive sensing'', \emph{42nd Annu. Conf. Inf. Sci. Syst.}, 2008, pp. 64--21.

\bibitem{c14} K. Knudson, R. Saab and R. Ward, ``One-bit compressive sensing with norm estimation'', \emph{IEEE Trans. Inf. Theory}, vol.62, no. 5, pp. 2784-2758, May. 2016.
\bibitem{c14a} S. Rangan, ``Generalized approximate message passing for estimation with random linear mixing", in \emph {Information Theory Proceedings (ISIT), 2011 IEEE International Symposium on}. IEEE, 2011, pp. 2168–2172.
\bibitem{c15} J. Wang, ``Support recovery with orthogonal matching pursuit in the presence of noise'', \emph{IEEE Trans. Signal Process.}, vol. 63, no. 21, pp. 5868--5877, Nov. 2015.
%\bibitem{c64} G. Reeves and M. Gastpar, ``The sampling rate-distortion tradeoff for sparsity pattern recovery in compressed sensing'', \emph{IEEE Trans. Inf. Theory.}, vol. 58, no. 5, pp. 3065--3092, May. 2012.
%\bibitem{c17} G. Reeves and M. Gastpar, ``Approximate sparsity pattern recovery: Information-theoretic lower bounds'', \emph { IEEE Trans. Inf. Theory.}, vol. 59, no. 6, pp. 3451--3465, Jun. 2013
\bibitem{c18} R. Blasco-Serrano, D. Zachariah, D. Sundman, R. Thobaben, and M. Skoglund, ``A measurement rate-MSE tradeoff in compressive sensing through partial support recovery'', \emph { IEEE Trans. Signal Process.}, vol. 62, no. 18, pp. 4643--4658, Sept. 2014.
\bibitem{c19} P. North and D. Needel, ``One-bit compressive sensing with partial support'', \emph{ 2015 IEEE 6th International Workshop on Computational Advances in Multi-Sensor Adaptive Processing (CAMSAP). IEEE}, 2015, pp. 349--352.
\bibitem{c20} L. Jacques, J. Laska, P. Boufounos and R. Baraniuk. ``Robust 1-Bit compressive sensing via binary stable embeddings of sparse vectors'', \emph{ IEEE Trans. Inf. Theory.}, vol. 59, no. 4, pp. 2082-2102, Apr. 2013.

\bibitem{c20_1} Y. Xie, Y. Eldar, A. Goldsmith, ``Reduced-dimension multiuser detection'', \emph{IEEE Trans. Inf. Theory}, vol. 59, no. 6, pp. 3858--3874, Jun. 2013.
\bibitem{c24} Y. Plan, R. Vershynin, ``Robust 1-bit compressed sensing and sparse logistic regression: A convex programming approach'',   \emph{IEEE Trans. Inf. Theory}, vol. 59, no. 1, pp. 482-494, Jan. 2013.
\bibitem{c25}  W. Tang, W. Xu, X. Zhang, et al, ``A low-cost channel feedback scheme in mmWave massive MIMO system". \emph{2017 3rd IEEE International Conference on Computer and Communications (ICCC)}, IEEE, 2017, pp. 89-93.

\end{thebibliography}
\end{document}